\documentclass[prb,showpacs,showkeys,letterpaper]{revtex4}
\usepackage{dcolumn}

\font\sc=cmcsc10

\def\unit#1{\,\hbox{#1}}
\def\m{\unit{m}}
\def\cm{\unit{cm}}
\def\s{\unit{s}}
\def\N{\unit{N}}
\def\kg{\unit{kg}}
\def\mps{\m\s^{-1}}
\def\mpss{\m\s^{-2}}
\def\x{\times}
\def\rhoair{\rho_{\rm air}}
\def\rhoball{\rho_{\rm ball}}
\def\ISBN{{\sc isbn}}

\begin{document}

\title{Introductory physics: The new scholasticism}

\author{Sanjoy Mahajan}
\affiliation{Physics Department, University of Cambridge, Cambridge CB3
  0HE, England}
 \email{sanjoy@mrao.cam.ac.uk}

\author{David W.~Hogg}
\affiliation{Physics Department, New York University, New York, NY
  10003, USA}
 \email{david.hogg@nyu.edu}

\date{29 March 2006}		

\begin{abstract}
\noindent Most introductory physics textbooks neglect air resistance
in situations where an astute student can observe that it dominates
the dynamics.  We give examples from many books.  Using dimensional
analysis we discuss how to estimate the relative importance of air
resistance and gravity.  The discussion can be used to mitigate the
baleful influence of these textbooks.  Incorrectly neglecting air
resistance is one of their many unphysical teachings.  Shouldn't a
physics textbook teach correct physics?
\end{abstract}

\maketitle

\section{The problem}
\noindent You teach introductory physics, say Newtonian mechanics.
You want students to use everyday experience to develop 
physical intuition.  On occasion you get a student who \emph{does} see
her or his experiences as a physics laboratory.  She avidly observes
sporting events, street life, and kitchen appliances for insight into
how macroscopic objects behave.  Then she encounters problem 4--34
from Halliday, Resnick \& Walker (see the list in
Table~\ref{table:textbooks}): to analyze a golf drive ignoring air
resistance; {\it i.e.,} with the ball traveling on a parabolic
trajectory.

\begin{table*}
\caption{\label{table:textbooks}Textbook prices,
masses, and problematic air-resistance problems.}
\def\tablesec#1{\noalign{\bigskip\centerline{\sc #1}\medskip}}
\def\0{\openup0.3\jot}
\def\1#1{\vtop{\0\hsize=0.3\hsize \raggedright #1}}
\def\2#1{\vtop{\0\hsize=0.25\hsize \raggedright #1}}
\def\3#1{\vtop{\0\hsize=0.2\hsize \raggedright #1}}
\def\4{\noalign{\vskip4pt}}
\begin{ruledtabular}
\begin{tabular}{llrrc}
\noalign{\smallskip}
\hfil \it Author/Title&\hfil \it Edition&\hfil \hbox{\it Mass}&
\hbox{\it Price\/}\rlap{\footnote{Mass and price from \url{amazon.com}
on 2006-03-28 using the {\tt isbn2info.py} script in the 
physics/0412107 source.}}&\hbox{\it Problematic problems}\\
\noalign{\smallskip}\hline
\tablesec{One of several volumes}
\1{D.~Halliday, R.~Resnick \& J.~Walker, {\it Fundamentals of
Physics}\/\footnote{At least Halliday {\it et al.} mention $v^2$ drag.}}&
\2{7th ed., volume 1, John Wiley, 2005, \ISBN{} 0-471-42959-7}&
3.4\,\rlap{lb}&
\hbox{US\$}93.95&
\3{4--34, 4--26 (tennis), 4--28 (soccer), and 4--37 (baseball)}\\ \4
\1{R.~Resnick, D.~Halliday \& K.~S.~Krane, {\it
Physics\/}\footnote{describes the drag force on a basketball or a
skydiver as proportional to $v$ rather than $v^2$ (p. 72)!}}&
\2{5th ed.\ volume 1, John Wiley, 2002, \ISBN{} 0-471-32057-9}&
3.4&
93.95\\ \4
\1{D.~C.~Giancoli, {\it Physics for Scientists \& Engineers}}&
\2{3rd ed., volume 1, Prentice Hall, 2000, \ISBN{} 0-13-021518-X}&
3.1&
102.67&
\3{3--22 (skiing), 3--31 (football), and 3--82 (baseball)}\\ \4
\1{A.~F.~Rex \&\ M.~Jackson, {\it Integrated Physics and Calculus}}&
\2{volume 1, Addison Wesley, 2000, \ISBN{}
0-201-47396-8}&
2.6&
73.00&
\3{4--21 through 4--24}\\ \4
\1{P.~A.~Tipler, {\it Physics for Scientists and Engineers}}&
\2{4th ed., volume 1, W.~H.~Freeman, 1999, \ISBN{} 1-57259-491-8}&
3.6&
84.95&
\3{3--88 (baseball)}\\
\tablesec{Complete in one volume}
\1{K.~Cummings, P.~W.~Laws, E.~F.~Redish \&
P.~J.~Cooney, {\it Understanding Physics\/}\footnote{discusses air
 resistance but not in relation to the relevant
problems, where it merely mentions that `In some of the problems,
exclusion of the effects of the air is unwarranted but helps simplify
the calculations' (p. 131).}}&
\2{Wiley, 2004, \ISBN{} 0-471-37099-1}&
6.4&
138.95&
\3{5--8 (golf), 5--15 (football), 5--18 (soccer), and 5--23 (baseball)}\\ \4
\1{P.~M.~Fishbane, S.~G.~Gasiorowicz \& S.~T.~Thornton, {\it Physics for
Scientists and Engineers with Modern Physics}}&
\2{3rd ed., Pearson Prentice Hall, 2005, \ISBN{} 0-13-035299-3}&
6.7&
160.00&
\3{3--44 (football) and 3--71 (golf)}\\ \4
\1{D.~C.~Giancoli, {\it Physics:\ Principles with Applications}}&
\2{6th ed., Prentice Hall, 2005, \ISBN{} 0-13-060620-0}&
5.2&
146.67&
\3{3--22 (football), 3--29 (football), 3--62 (baseball), and 3--68 (skiing)}\\ \4
\1{R.~D.~Knight, {\it Physics for Scientists and Engineers with Modern
Physics: A Strategic Approach}}&
\2{Addison--Wesley, 2004, \ISBN{} 0-8053-8685-8}&
9.9&
160.00&
\3{6--29 (tennis) and 6--38 (football)}\\ \4
\1{R.~A.~Serway \& J.~W.~Jewett Jr., {\it Principles of Physics: A
Calculus-Based Text}}&
\2{3rd ed., Brooks/Cole, 2002, 
\ISBN{} 0-03-027157-6}&
6.0&
148.95\rlap{\footnote{from price sticker on the back (no Amazon
list price).}}&
\3{3--15 (football) and 3--45 (baseball)}\\ \4
\1{J.~S.~Walker, {\it Physics}}&
\2{2nd ed., Pearson 
Prentice Hall, 2004, \ISBN{} 0-13-101416-1}&
5.7&
146.67&
\3{4--26 (soccer), 4--31 (golf),
4--33 (football),
4--42 (golf), 4--46 (soccer), and 4--48 (golf)}\\ \4
\1{H.~D.~Young \& R.~A.~Freedman, {\it University Physics with Modern
Physics}}&
\2{11th ed., Addison Wesley, 2004, \ISBN{}
0-8053-8684-X}&
7.4&
160.00&
\3{3.15 (football), 3.17 (flare gun), 3.19
(baseball), and 3.85 (soccer)}\\ \4
\end{tabular}
\end{ruledtabular}
\end{table*}

The observant student knows---from exploring the world---that golf
drives rise quickly and almost straight and drop parabolically only
near the end.  This student learns that the world described by her
physics textbook is not the \emph{real} world and that careful
observation is irrelevant to physics.  

Contrast the experience of the curious student with that of a student
who parrots equations and regurgitates textbook paragraphs.  This
student is untroubled by the golf problem or its variants listed in
Table~\ref{table:textbooks} because he knows the easily memorized
`fact' that all trajectories are parabolae.  The parrot student
correctly answers this problem, confirming that physics
means memorizing disembodied mathematical facts.  Observant
student: 0; parrot student: 1.

Our objection is not only to the treatment of air resistance.  Most
textbooks leave many assumptions unjustified and unexamined: be they
massless, inextensible strings, frictionless pulleys, or pointlike
particles.  Air resistance, or its absence, is merely one glaring
example, which we now discuss in detail.


\section{Dimensional analysis}
You are a physicist.  You begin any solution by asking not what
equation matches but what physics is relevant and what approximations
to use.  Studying the flight of a golf ball, \emph{as a start}
you might neglect air resistance; then from
dimensions you could argue
that the velocity of a golf ball hit a distance $L$ (say,
$250\m$) is roughly
\begin{equation}
\label{eqn:gL}
v\sim\sqrt{gL}\sim(10\mpss\x250\m)^{1/2}\sim 50\mps.
\end{equation}
  
You check this assumption by comparing the gravitational force,
about $0.5\N$, to the drag force $f$ (from
air resistance).  By dimensional analysis again,
\begin{equation}
f\sim\rho Av^2,
\end{equation}
 where $\rho$ is the density of air
and $A$ is the cross-sectional area of the golf
ball.  The drag is due to the power fed into the
turbulent eddies that eventually become small eddies gobbled by
viscosity.  The analysis of the turbulence is difficult, if not
impossible, but dimensional analysis gives the correct form of the
drag force.

With this result and reasonable values $\rho\sim\sim1\kg\m^{-3}$
and $A\sim 10^{-3}\m^2$, the drag force is
\begin{equation}
\label{eqn:drag-estimate}
f\sim 1\kg\m^{-3}\x10^{-3}\m^2\x(50\mps)^2\sim
2.5\N.
\end{equation}
Since the gravitational force is roughly $0.5\N$,
the drag force is \emph{5~times the gravitational force!}  So
you reject the original assumption of no drag and numerically
integrate to approximate the trajectory (easy with spreadsheets used
by students in other contexts).  An early conclusion of your
approach---the physical approach---is that the golf ball must be hit
much faster than $50\mps$, which is the speed it would have to be hit if there
were no drag.  Therefore the drag is even larger than the estimate
in (\ref{eqn:drag-estimate}).


\section{Dimensionless ratios}
As a physicist, you also know about the physical insight provided by
dimensionless ratios.  The interesting ratio here is of drag force to
gravitational force:
\begin{equation}
{\hbox{drag force}\over\hbox{gravitational force}}\sim
{\rhoair A v^2\over mg}.
\end{equation}

If $d$ is the diameter of the golf ball,
then its cross-sectional area is $A\sim d^2$ and 
its mass is $m\sim\rhoball d^3$.   Furthermore 
the $v^2$ in the drag force is roughly $gL$, from the estimate (\ref{eqn:gL}).
So, after dividing out the common factor of $g$,
\begin{equation}
\label{eqn:ratio}
{\hbox{drag force}\over\hbox{gravitational force}}\sim
{\rhoair Ld^2\over \rhoball d^3}.
\end{equation}
The denominator 
is mass of the golf ball.  What about the numerator?
Since $Ld^2$ is roughly the volume of air that the golf ball sweeps out,
the numerator is
\begin{equation}
\rhoair Ld^2 \sim \rhoair\x\hbox{volume swept out},
\end{equation}
which is the mass of air that the golf ball sweeps out.  So
\begin{equation}
{\hbox{drag force}\over\hbox{gravitational force}}\sim
{\hbox{mass of air golf ball sweeps out}\over\hbox{mass of golf ball}},
\end{equation}
or in English:
\begin{quotation}
{\bf Air resistance is significant if the golf ball sweeps out a mass
comparable to itself.}
\end{quotation}


In the ratio (\ref{eqn:ratio}),
two powers of the diameter $d$ divide out:
\begin{equation}
{\hbox{drag force}\over\hbox{gravitational force}}\sim
{\rhoair\over \rhoball}\x{L\over d}.
\end{equation}
The density ratio is $\sim10^{-3}$:
Golf balls roughly float on water so
$\rhoball\sim10^3\kg\m^{-3}$.  The length ratio is
roughly
\begin{equation}
{L\over d}\sim{250\m\over4\cm}\sim6000,
\end{equation}
where $d\sim4\cm$ is a typical golf-ball diameter and $L\sim250\m$ is
the range used in (\ref{eqn:gL}).  The product of the density and length
ratios is~6, which echoes the ratio of~5 between drag and gravity
estimated after (\ref{eqn:drag-estimate}).

This argument---a quantitative argument about the relative importance
of gravity and air resistance---uses neither advanced mathematics nor
abstruse concepts.  Yet it is sophisticated and leads to a general
insight connecting the importance of air resistance and the ratio of
displaced masses.  It is sophisticated because it uses the physicist's
special tools: estimation and dimensional analysis.  This argument
appears in \emph{no} introductory text.  Why not?  Not because
students find it difficult, for it applies simple principles and
mathematics to familiar objects.  Not because such arguments have
failed, for they have not even been tried.  Perhaps it does not appear
in introductory textbooks because it refutes so many standard
problems.

Of the textbooks in Table~\ref{table:textbooks} that come in a single
volume (second portion of the table), the average price is \$152 and
the average weight is 6.8~pounds.  We should not teach from, take
problems from, or ask students to buy obese, expensive books filled
with bogus physical analyses of easily observed phenomena.  If you
agree with this principle, then you will not voluntarily assign or use
the textbooks in Table~\ref{table:textbooks} nor others that
incorrectly analyze golf shots, baseball hits, tennis serves, downhill
skiing, or any phenomena where air resistance is important.  \emph{The
requirement that physics textbooks teach correct physics excludes most
books on the market.}

One book treats air resistance correctly: {\it Matter \&\ Interactions
I:\ Modern Mechanics\/} \cite[pp.~180--188]{Chabay-Sherwood:2002}.  In
problem 5.3, students numerically integrate the
trajectory of a baseball with and without air resistance.  The authors
point out that `the effect is surprisingly large---about a factor of
2!'  The problem even compares the effect of air resistance at high
altitude (Denver) with its effect at sea level.  This book and the
companion volume on electricity and magnetism, which have been
reviewed by Titus,\cite{Titus:2006}
are highly recommended to teachers who want to
return physics to introductory physics.

\section{Calculus textbooks}
Physics textbooks often ignore air resistance.  Articles on teaching
often perpetuate an alternative misconception: that drag is linear in
$v$.  For example, Warburton and Wang \cite{Warburton-Wang:2004}
analyze the large-velocity limit saying: `For simplicity, we will
assume linear air resistance\dots'!  The mathematics textbooks,
perhaps where many of us first saw analyses of air resistance, are
hardly role models.  One author (SM) was mistreated by the following
argument in high-school calculus.  The author wrote the drag force
as a Taylor series in $v$:
\begin{equation}
f = c_1v + c_2v^2 + c_3v^3 + \cdots.
\end{equation}
Following the usual practice with $\sin x\approx x - x^3/3+\cdots$, 
he argued that the term with the lowest power of
$v$, in other words $c_1v$, is the dominant term.  Therefore
\begin{equation}f\propto v\qquad\hbox{(approximately)}.\end{equation}
The gentle reader will enjoy finding the flaw in this argument.

In a current calculus textbook,\cite{Thomas:2003}
the introduction offers heartening news (p.~viii) `\dots we have
not compromised our belief that the fundamental goal of a calculus book
is to prepare students to enter the scientific community.'  
Sadly, the section on First Order Differential Equations (the usual location
of this unphysical result) contains the ominous subhead `Resistance
proportional to velocity' (p.~534).  Therein the reader learns:
\begin{quotation}
In some cases it makes sense to assume that, other forces being
absent, the resistance encountered by a moving object, like a car
coasting to a stop, is proportional to the object's velocity.  The
slower the object moves, the less its forward progress is resisted by
the air through which it passes.
\end{quotation}
A car coasting to a stop!  Its Reynolds number is
\begin{equation}{\rm Re}\sim {vl\over\nu},\end{equation}
where $\nu1.5\x10^{-5}\m^2\s^{-1}$
is the kinematic viscosity of air, and $l$ is a typical
length scale for a car, perhaps $2\m$.
At a typical speed of $v\sim20\mps$, the Reynolds number is
\begin{equation}{\rm Re}\sim {20\mps\x2\m\over1.5\x10^{-5}\m^2\s^{-1}}\sim 3\x10^6.\end{equation}
Drag switches from $v^2$ (turbulent) to $v$ (Stokes) drag
at a Reynolds number of say ${\rm Re}\sim 3$ (see Tritton's excellent
textbook\cite[pp.~32--34]{Tritton:1988}), which occurs when $v$ falls
by a factor of $10^6$ so when
$v\sim2\x10^{-5}\m\s^{-1}$.  The car has certainly coasted to a stop!

Digging the hole deeper, the section offers an example (p.~536): 
\begin{quotation}
Example 8.  For a 192-lb ice skater, the $k$ in [$F_{\rm drag}=kv$] is
about 1/3 slug/sec and $m=192/32 = 6$ slugs.  How long will it take
the skater to coast from 11 ft/sec (7.5 mph) to 1 ft/sec?  How far
will the skater coast before coming to a complete stop?
\end{quotation}
Leaving aside the obscure units, the authors take formulae derived
from bogus physics and ask students to insert alleged constants into
the resulting bogus formulae!


\section{The desert}
We live and teach in the desert of the real
\cite[p.~1]{Baudrillard:1995}.  Our betters may force us to
assign a book in Table~\ref{table:textbooks}.  We can mitigate its
harm by teaching students how to make quantitative checks on the
approximations and assumptions and by class discussion on dubious
problems.

Great value still lies in explaining and analyzing trajectories
neglecting air resistance, such as the trajectory of a baseball or
golf ball when tossed gently from hand to hand or of a solid, dense
metal ball falling a few meters.  The golf drive can be analyzed
neglecting air resistance so long as the problem's unreality is
emphasized and physically reasonable approaches follow.  Our
recommendation need not make introductory physics more difficult.  We
just want it to become more \emph{physical}.

If a student can tell you that an object with constant acceleration
moves on a parabolic trajectory, then the student has learned or
memorized a mathematical fact.  If a student can tell you when and why
constant acceleration usefully approximates a real system, then the
student has learned physics and understands important patterns in our
Universe.  Physics studies the world, not just mathematical
relationships.  By that standard, introductory `physics' has become as
remote from students' reality as have medieval disputations about the
size of angels.

\bigskip
\begin{acknowledgments}
We thank Mike Blanton, David Goodstein, Bruce Sherwood, and Aimee
Terosky for helpful comments and discussion.
\end{acknowledgments}

\bibliography{polemic}

\end{document}